\documentclass[onecolumn]{IEEEtran}
\usepackage{amsmath}
\usepackage{theorem}
\usepackage{amsmath}
\usepackage{mathrsfs}
\usepackage{accents}
\usepackage{setspace}
\usepackage{url}
\usepackage{algorithm}
\usepackage{algorithmic}
\usepackage{epsfig}
\usepackage{amsfonts}
\usepackage{amssymb}
\usepackage{mathrsfs}
\usepackage{euscript}
\usepackage{graphicx}
\usepackage{multirow}
\usepackage{cite}

\usepackage[T1]{fontenc}
\usepackage[USenglish,UKenglish,french,spanish,italian]{babel}
\usepackage[nodayofweek,level]{datetime}
\newcommand{\mydate}{\formatdate{2}{12}{2016}}

\graphicspath{{./}{figures/}}

\ifCLASSINFOpdf
\else
\fi
\begin{document}

\begin{titlepage}

\begin{tabular}{l        r}

\includegraphics[bb=20bp 00bp 500bp 450bp,clip,scale=0.3]{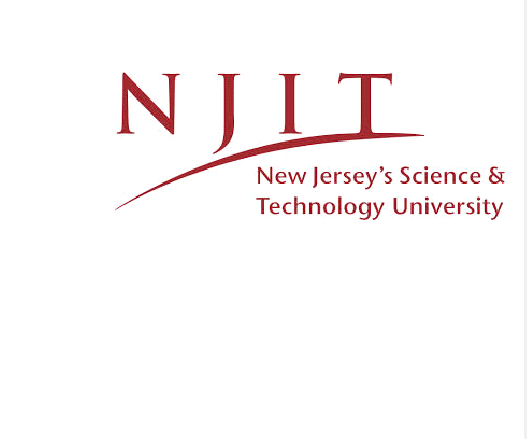} \hspace{6cm} & \includegraphics[bb=0bp -200bp 500bp 550bp,clip,scale=0.2]{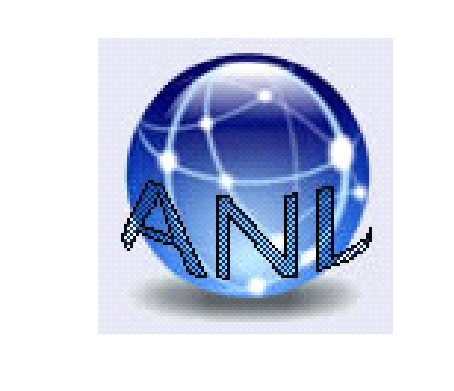}

\end{tabular}

\begin{center}

\textsc{\LARGE On The Fundamental Energy Tradeoffs of Geographical Load Balancing}\\[1.5cm]

{\Large \textsc{Abbas Kiani}}\\ 
{\Large \textsc{Nirwan Ansari}}\\ 
[2cm]

{}
{\textsc{TR-ANL-2016-003}\\
\selectlanguage{USenglish}
\large \mydate} \\[3cm]

{\textsc{Advanced Networking Laboratory}}\\
{\textsc{Department of Electrical and Computer Engineering}}\\
{\textsc{New Jersy Institute of Technology}}\\[1.5cm]
\vfill

\end{center}

\end{titlepage}


\selectlanguage{USenglish}
\begin{spacing}{2}
\begin{abstract}
Geographical load balancing can optimize the utilization of green energy and
the cost of electricity by taking the advantages of green and price diversities at geographical dispersed data centers.
However, higher green energy utilization or lower electricity cost may actually increase the total
energy consumption, and is not necessarily the best option. The achievable energy tradeoffs can be captured by taking into consideration of  a defined service efficiency parameter for geo-dispersed data centers.
\end{abstract}

\section{Introduction}\label{sec:Introduction}
The demand for online services including web search, online gaming, distributed file systems such as Google File System (GFS), and distributed Storage System such as BigTable and MapReduce is growing exponentially. This explosion of demand for online services has led to a multitude of challenges in  Data Center Networks (DCNs) from DCN architecture design, congestion notification, TCP Incast, virtual machine migration, to routing in DCNs~\cite{zhang2013architecture}.

Most importantly, data centers electric power usage is growing at a rapid pace. In 2013, U.S. data centers consumed an estimated 91 billion kilowatt-hours of electricity, and as the fastest growing consumer of electricity, they are estimated to consume roughly 140 billion kilowatt-hours in 2020 which will cost \$13 billion in electricity bill and emit 100 million metric tons of carbon pollution~\cite{NRDC}.
This huge average annual electricity consumption is due not only to the the continuing explosion of Internet traffic but also to the gravity of preparing DCNs as a scalable and reliable computing infrastructure.
Online services run on hundreds of thousands of servers spread across server farms provisioned for the peak load. In fact,
to assure the user demands satisfaction, the servers run 24/7 and in vast underutilization the majority of the time.
To put this in perspective, the total power consumption at a data center includes the Base Load and Proportional Load. The base load indicates the power consumption even when some of the turned on servers are idle. On the other hand, the proportional load is the extra power consumption which is proportional to the CPU utilization of the servers and accordingly to the load.
 Therefore, even being idle, servers draw the base load power,
thus incurring a substantial amount of annual energy use. However, in the past few years, more server capacities have been virtualized to facilitate multiple Virtual Machines (VMs) being run on a single Physical Machine (PM).

Complying with all of our online activities but limiting the increasing energy demand in an environmentally friendly manner calls for innovations across different disciplines.
Recently, a great deal of research has been done to cut the data center's power consumption and accordingly the cost of electricity.
A great part of the studies mainly aims at proposing new power management techniques by investigating the CPU and memory power consumption of the servers. For examples, Dynamic Voltage/Frequency Scale (DVFS) schemes like~\cite{DVFS1} have been deployed to reduce the CPU power and new techniques such as~\cite{memorypower2} have been proposed to adjust the power states of the memory devices in order to dynamically limit memory power consumption. However, the data center operators prefer to maintain a high level of reliability and uptime with their less expensive inefficient facilities rather than to install energy efficient devices at the cost of higher upfront price~\cite{NRDC}.

Opportunities to improve the data centers energy efficiency is not limited to the improvements in computing components. The energy consumption break down of data centers shows that a course of action is required to improve the energy consumption at other components like network equipment, electrical power delivery and conversion, cooling, and lighting.
To this end, Power Usage Effectiveness (PUE) metric has been commonly adopted as a measure of data centers efficiency, and is defined as the ratio of the total energy consumed by the data center to that consumed by the Information Technology (IT) equipment (EPA report on server and data center energy efficiency, Final Report to Congress, Aug. 2007).
Power delivery and cooling efficiency has been the subject of interest of many recent research papers, and a large number of studies have aimed at innovating networking components and topologies to shave the power consumed by the IT network.

Another approach which addresses the energy consumption in all components is referred to as green data centers.
The concept not only tries to cut down the electricity consumption and its cost but also integrates renewable energy resources such as solar panels and wind farms into data centers, thereby promoting sustainability and green energy. The data center operators can assess the sustainability of their data centers using the Carbon Usage Effectiveness (CUE) metric along with PUE. CUE is defined as the ratio of the total $CO_2$ emissions caused by the total data center energy consumption to that by the IT equipment energy consumption. CUE has the ideal value of 0.0 which indicates no carbon use is associated with the data center operations~\cite{belady2010carbon}.

Shaving the energy consumption and its cost via load shedding and load shifting (\cite{wierman2014opportunities} and references therein) is another approach.
Load shedding is associated with QoS degradation where data centers based on the Service Level Agreements (SLAs) decide to serve some types of the workload less effectively by utilizing less energy. On the other hand, load shifting algorithms investigate the possibility of shifting the load in time to run when for example cheaper electricity is available.
Moreover, the effectiveness of the geographical load balancing on the energy costs and utilization of more renewable energy has been demonstrated in some studies. In the so called Geographical Load Balancing (GLB), the workload is distributed among Internet scale data centers spread across geographical diversity~\cite{liu2015greening}. In the following sections, we will investigate the opportunities and challenges of geographical load balancing.

\section{Green and Economic Geographical Load Balancing}

\begin{figure}
\center
\includegraphics[width=13cm,height=10cm]{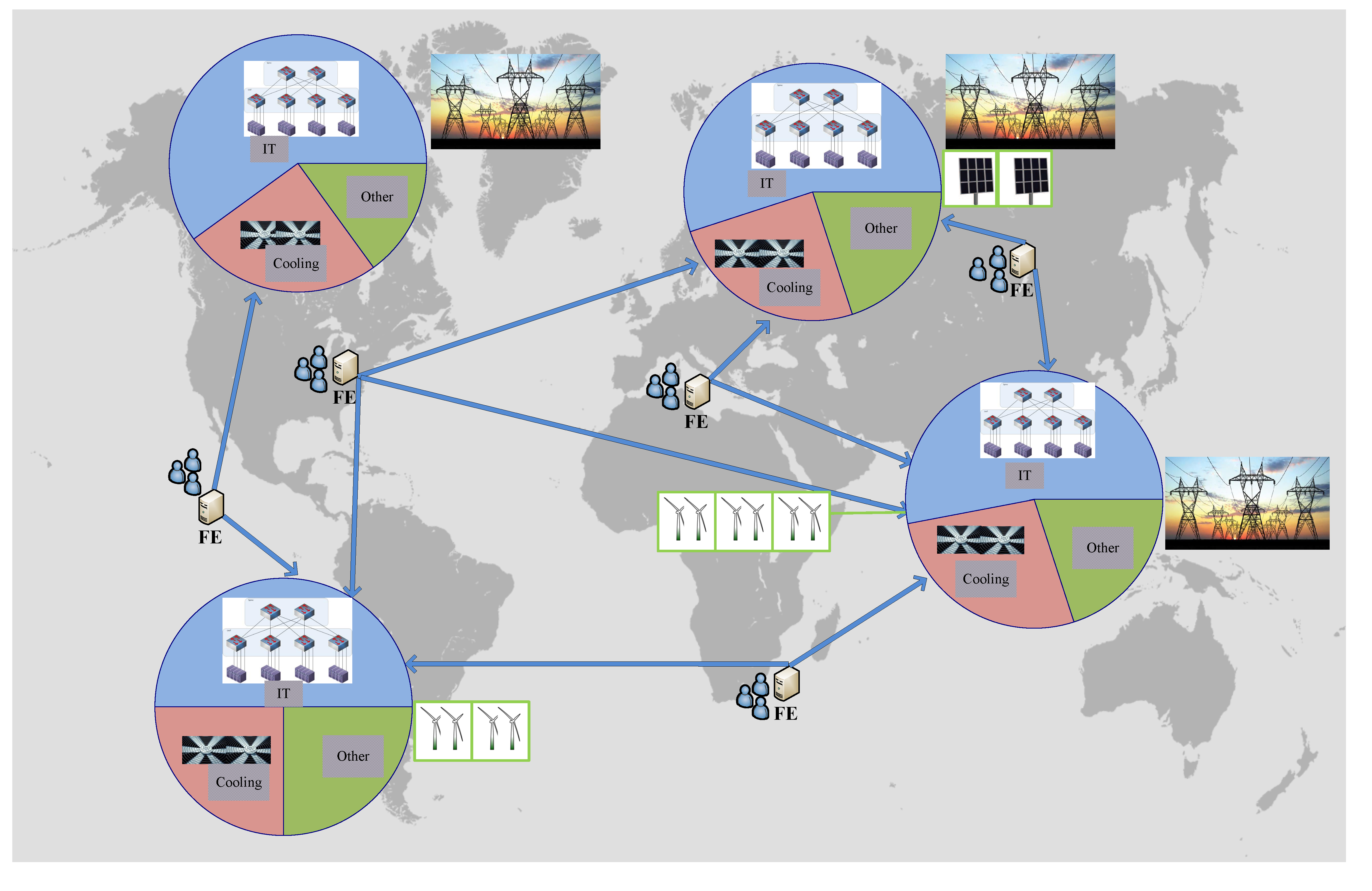}
\caption{Geographical dispersed data centers.}
\label{fig:1}
\end{figure}

The Internet scale powerful data centers are few because of the scale and cost of the deployment and operation. These few numbers of data centers are generally dispersed in different geographical regions. The main goal of deploying such geo-dispersed data centers is not only to provide redundancy, scalability and high availability but also to more efficiently employ global resources such as utilizing price-diversity in electricity markets or utilizing locational diversity in renewable power generation~\cite{liu2015greening,kiani2015towards} (see Fig.~\ref{fig:1}).
Therefore, the powerful data centers are generally deployed far away from a large majority of users. To this end, Front-End (FE) servers are co-located with users. Each FE server receives requests from its nearby users and distribute the requests to the back-end servers at geo-dispersed data centers.
In fact, each FE server functions as a workload distribution center that manages the workload by distributing the user requests to the appropriate data centers.

The selection of the appropriate data centers can be based on different parameters like server or content availability, the network distance between FE and data center, the efficiency of the data centers, the cost of the electricity, and availability of the renewable energy.
Therefore, different workload distribution strategies can be adopted at each FE by considering different objectives like maximizing green energy utilization, minimizing the cost of electricity or maximizing the profit gained by running data center networks.
On the other hand, each service request has to be handled within a deadline determined by
the Service Level Agreement (SLA). Different parameters like the throughput of the connection between users and FE server, FE server and back end servers at the data center, the queuing and processing delay at the data center are contributing to the end-to-end delay of a service request~\cite{chen2011characterizing}. The QoS at a data center is generally enured by implying an upper bound on the
queuing delay at the data center which has been commonly modeled as M/GI/1 Processor Sharing (PS) queue or M/M/1 queue~\cite{liu2015greening}.

\subsection{Green-Diversity}

To benefit from the energy efficiency and sustainability
advantages of greening, data centers have been recently integrated with a green power source such as wind turbine or solar panel.
There are three different ways to green a data center.
The first approach, called behind the meter, is to install renewable power generators at the data center location.
In this case, the data center operator can own the power generation system itself or a third party can install the system and sell the generated power to the data center. However, the most efficient location to build a renewable power source is not always the same as the best location to build an efficient data center. Therefore, data center operators such as Google choose to either purchase Renewable
Energy Certificates (RECs) or make Power Purchase Agreements (PPAs) to procure both power and RECs~\cite{google}.

To maximize green energy utilization, one FE server can manage the distribution of its incoming workload to different data centers based on the availability of green energy. The available green energy at a data center can be determined by the green energy generation or storage at the data center. The generated on-site green energy at a data center can be predicted
by taking into account of weather dependency of
green energy. Specifically, when the renewable generator is a wind turbine, the prediction can rely on the foremost forecasting
techniques which are based on Numeric Weather Prediction (NWP) of wind speed and power~\cite{soman2010review}. The prediction may
include Very-Short Term Forecasting, Short Term Forecasting, Medium Term Forecasting and Long Term Forecasting techniques.
If the case is solar generation, machine learning based prediction techniques can be employed.
In the case of purchased green energy, although it is not possible to track the flows of green energy from grid, green energy generation can be estimated via data center's RECs.
Moreover, when extra green energy is available, each data center can store green energy at energy storage devices and draw the energy from the storage device later.

\subsection{Price-Diversity}

While the data centers operate 24/7, the green energy is not a constant available resource to power them. Therefore, the data centers have to be connected to on-grid brown energy.
In this case, we should note that the brown energy is procured in deregulated electricity markets.

Unlike the regulated electricity markets, in deregulated electricity markets such as
day-ahead and real-time markets, the electricity prices vary during the day. The final prices are set based on the bidding process between the energy suppliers and consumers. Some studies also suggest that the data centers can participate in the bidding process and procure the electricity directly from the wholesale market~\cite{ghamkhari2013profit2}. However, the prices are not known to the data centers until the operating time.
For example, the day-ahead prices are usually revealed several hours up to one day
in advance while the real-time prices are known only a few minutes in advance. Therefore, the electricity price forecasting methods have to be employed when participating in biding process.

GLB can be considered as an opportunity to reduce the cost of electricity by utilizing electricity price diversity at different locations. In other words, in order to minimize the electricity cost, each FE server can mange the workload by sending the requests to the data center locations with cheaper price of electricity.

\subsection{Green Versus Brown}

Green and price diversities are considered as an opportunity to design a green and low cost GLB approach that not only can maximize the utilization of green energy but also minimize the cost of electricity.
However, due to the different costs and different environmental
impacts of the renewable energy and brown energy, such a GLB approach should tap on the merits of the separation
of green energy utilization maximization and brown energy cost minimization problems. That is, the concept of decomposing the workload to the workloads served by green and brown energy~\cite{kiani2015towards}. In other words, the notion of green workload and green
service rate, versus brown workload and brown service rate,
respectively, to facilitate the separation of green energy utilization
maximization and brown energy cost minimization problems.

\begin{figure}
\center
\includegraphics[width=13cm,height=10cm]{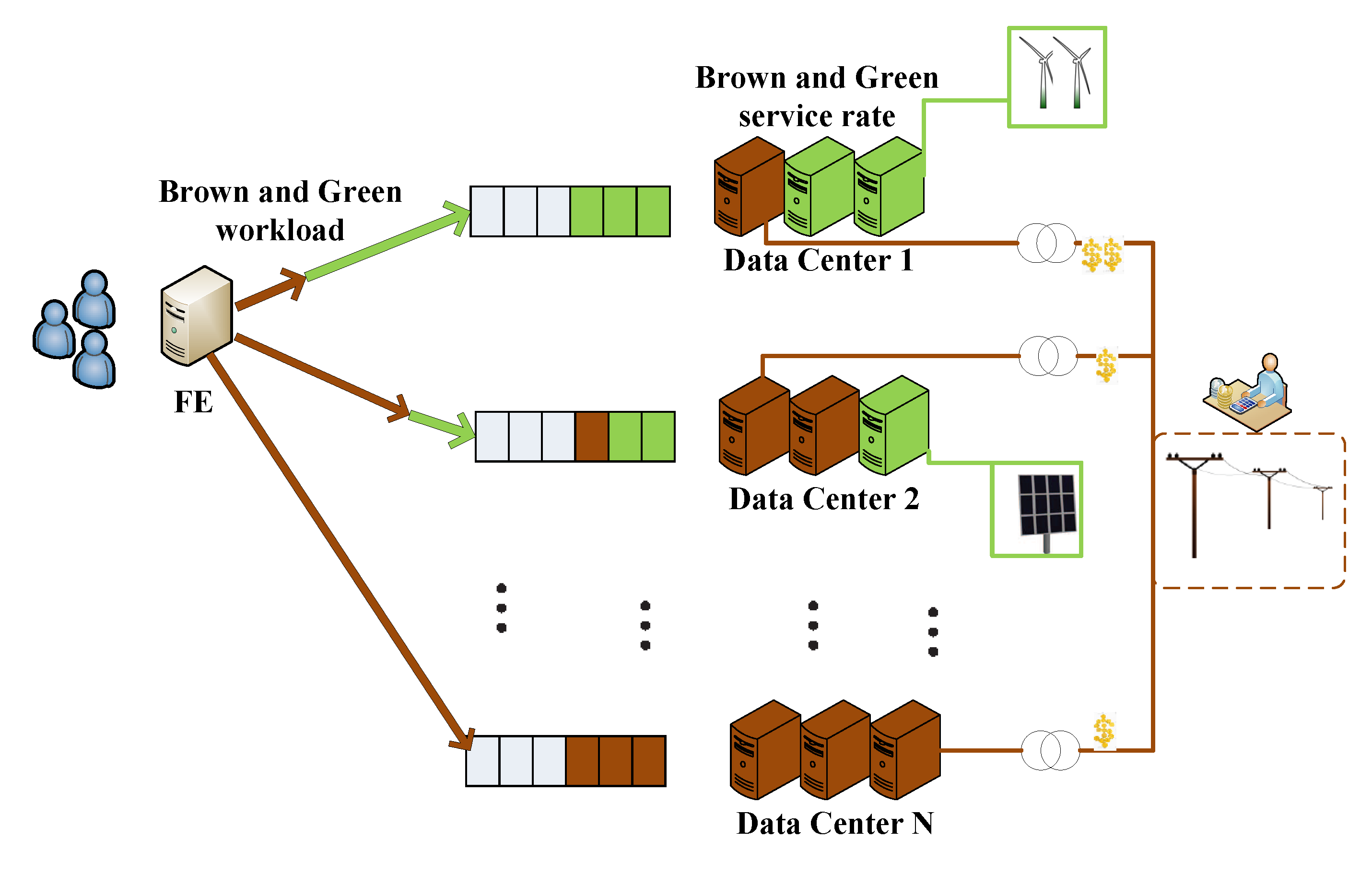}
\caption{Green versus brown.}
\label{fig:2}
\end{figure}

The idea is to distinguish the servers at each data center based on the energy which is utilized to power them (see Fig.~\ref{fig:2}). In fact,
some servers are turned on and powered by the available green energy (green servers)
and the others if needed by purchasing brown energy (brown servers). Therefore, the distinction
between green and brown workloads is made mainly based on the server which is utilized to serve
the workload. In specific, the workload served by a green server is defined as the green workload
and similarly the workload served by a brown server is defined as the brown workload. Moreover, using this idea, we can tackle the shortcoming in some studies, which propose an integrated optimization framework but under the assumption that local renewable generation is always less than the local power consumption. In fact, using the green versus brown concept, each data center utilizes green energy as much as possible, and purchases brown energy only when the
green energy generation is not adequate to serve all incoming workloads.

In terms of the optimization models, the optimization framework for green workload allocation
has to maximize the utilization of green energy and the constraints are to limit the allocated green workload by the available green
resources. If the green energy is not adequate to satisfy the QoS requirements, another optimization framework is then designed to allocate the brown workload. While the objective of this optimization framework is to minimize the cost of electricity, the constraints are defined to enforce QoS requirements at each data center.

\begin{figure}
\center
\includegraphics[width=9cm,height=8cm]{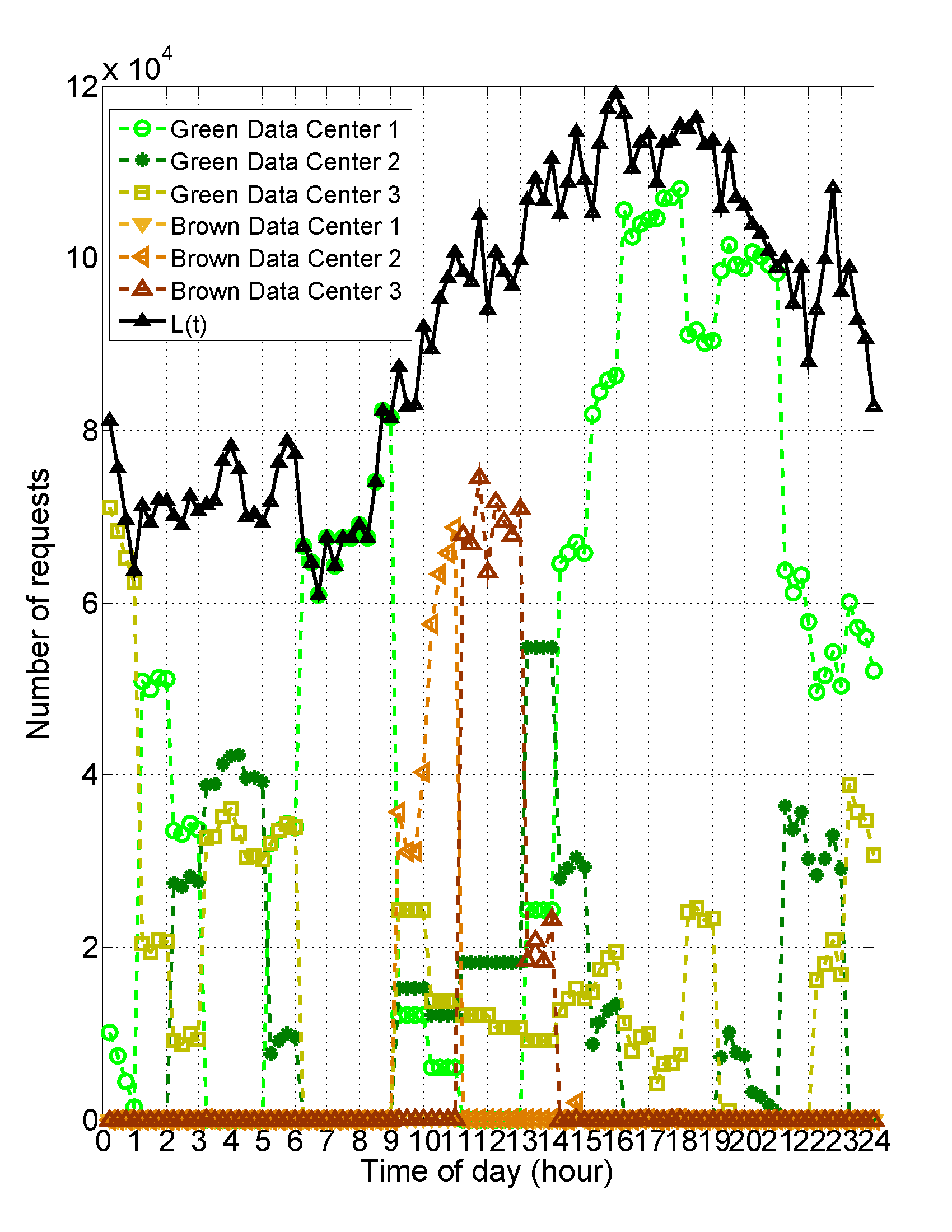}
\caption{Total incoming workload and allocated green and brown workloads to the data centers.}
\label{fig:3}
\end{figure}

\begin{figure}
\center
\includegraphics[width=9cm,height=8cm]{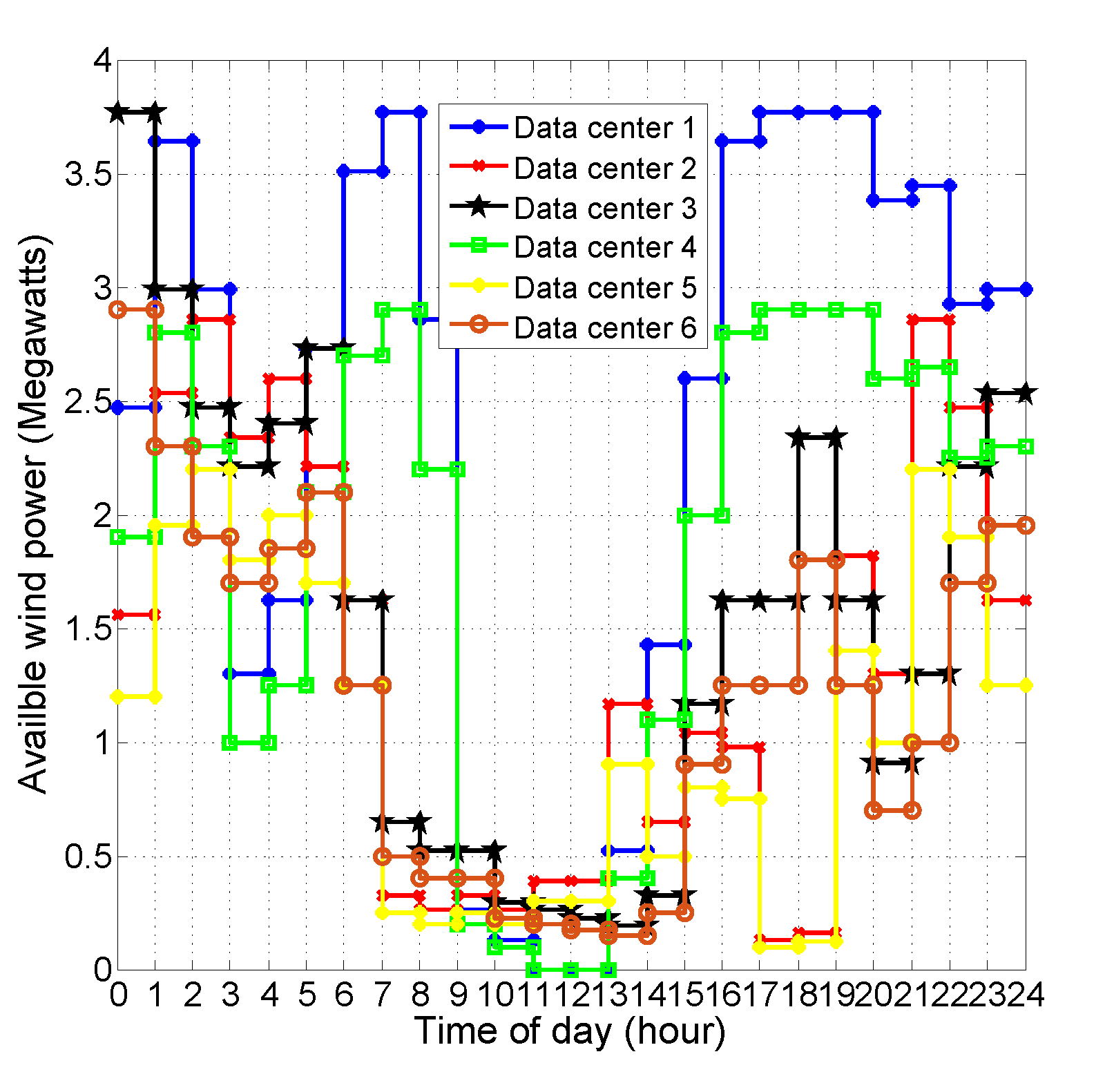}
\caption{Wind power generation.}
\label{fig:4}
\end{figure}

Fig.~\ref{fig:3} demonstrates the optimized allocated green and brown workload to $N=3$ data centers. The simulation data are based
on the trends of wind power at data centers 1,2 and 3 shown in Fig.~\ref{fig:4} and electricity price used in~\cite{kiani2015profit}. We simulated the total incoming workload by scaling up the trends of a sample day of the requests made to the 1998 World Cup web site which is also depicted in Fig.~\ref{fig:3}. In the simulations, we consider a time slotted system such that our optimization is applied in each time slot, and the QoS at each data center is enforced by maintaining an upper bound
on our estimation of data center's queue length at the next time slot.

For example, the trend of wind power indicates
that after hour 14, more of the green workload is assigned
to data center 1 where the highest wind power is available.
However, from hours 10 to 13, the available wind power at
data center 1 is lower than the other data centers, and thus less
of the green workload is allocated to this data center. Moreover, as
shown in Fig.~\ref{fig:3}, from hours 9 to 11, all of the left
over of the requests (brown workload) are allocated to data center 2 where
the price of electricity is the lowest. Note that before hour 8 and from hours
15 to 24, the available wind power is adequate to serve all the
service requests, and the brown workload
is almost not allocated to the data centers. In other words, in these hours, the available wind power is the key decision
factor to allocate workloads among the data centers.

\section{Fundamental Tradeoffs}

Most of the proposed GLB strategies aim at reducing the energy cost or brown energy consumption via distributing the requests to the locations with cheaper price of electricity or higher renewable energy generation. However, such strategies may increase the total power consumption due to the fact that different data centers have different servers with different service capabilities, and also a request sent to different data centers experiences different network delays. Therefore, consuming the same or even more amount of energy at one data center may handle less number of requests than another data center. In other words, the idea of sending a request to another data center with higher network delay or less service capability only in order to benefit from cheaper electricity or utilize more renewable energy may lead to a significant increase in the total power consumption.

The extra green energy generation at a data center can be injected into the power grid, and the data center can receive compensation for the injected power. In the case of electricity, the cheap electricity at a data center can be stored at energy storage devices to be utilized later when the electricity becomes more expensive. Therefore, the more green energy utilization or the cheaper electricity at the expense of increasing the total energy consumption is not necessarily the best option.

\subsection{Information Flow Graph Based Model}

\begin{figure}
\center
\includegraphics[width=9cm,height=9cm]{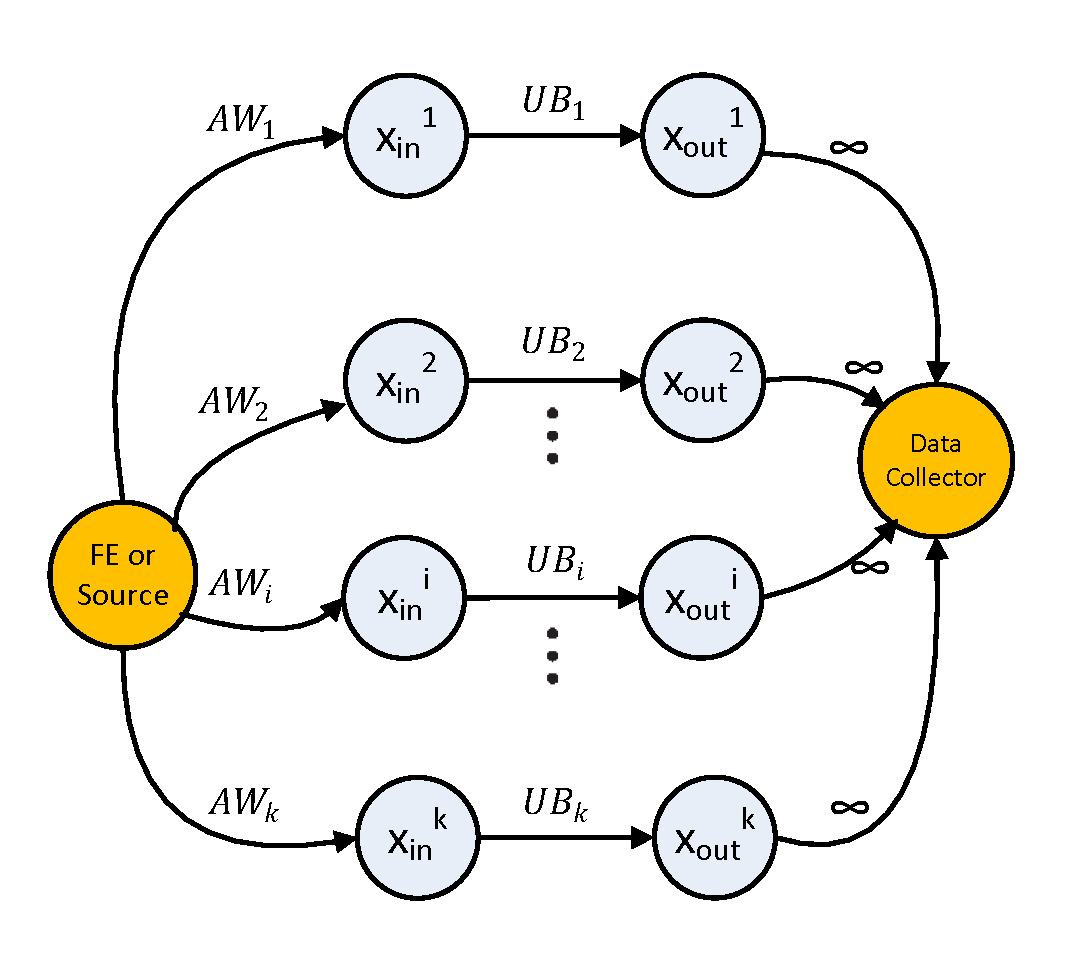}
\caption{Information flow graph based model for geo-dispersed data centers.}
\label{fig:5}
\end{figure}

To find the achievable tradeoffs between total power consumption and green energy utilization, we propose to model geo-dispersed data centers with an information flow graph~\cite{kiani2016tradeoff} (see Fig.~\ref{fig:5}). Note that this idea may be adopted to capture the achievable tradeoffs between total power consumption and the cost of electricity.

The information flow graph is a directed acyclic graph which includes three types of nodes: (i) a single source node (S), (ii) some intermediate nodes, and (iii) data collector nodes~\cite{dimakis2010network}. As depicted in Fig.~\ref{fig:5}, FE can be thought as the source node which is the source of original requests (FE node). Here, we only consider one FE but the proposed model can be extended to multiple FEs as the geographically concentrated sources of requests. The intermediate nodes are also data centers, and the data collector node can correspond to the users that receive processed requests.

The information flow graph, which models the geo-dispersed data centers, varies across time. At any given time, each node in the graph is either active or inactive. At the initial time of each time slot, the FE node as the only active node contacts all data center nodes and
sorts them based on a defined service efficiency parameter.
We define the service efficiency parameter based on an M/GI/1 Processor Sharing (PS) queue analysis by taking into consideration of the network delay between FE and data centers.
In other words, the allocated requests to a data center are first placed in a queue before they can be processed by any available server. We model the queue at each data center as an M/GI/1 PS queue which has been commonly
adopted in modeling the waiting time of the requests at a data center in many studies like~\cite{liu2015greening}. The M/GI/1 PS queue is a single server queue with Poisson arrivals in which the service times form a sequence of i.i.d random variables with an arbitrary and general distribution function. Under PS, the processor is shared fairly
among all jobs in the system.
Therefore, the queuing delay at a data center can be computed as a function of the allocated requests to the data center, the service rate of a single server at the data center, and the total number of the turned on servers at the data center.
The total number of turned on servers at a data center also depends on
the power consumption of the data center, the average peak power of a turned on server in handling a service request, and the Power Usage Effectiveness (PUE) of the data center~\cite{kiani2015towards,ghamkhari2013profit2}.

 The FE node connects to a set of data center nodes with capacities of the edges equal to the Allocated Workloads (AW) to these nodes.
To satisfy the QoS requirements, the allocated workload to a data center has to be upper bounded. In fact, the queueing delay for each service request should be limited by a given deadline determined by the Service Level Agreement (SLA); this constraint is then translated to an Upper Bound (UB) for each data center.
The value of this upper bound is varied at different data centers due to different service efficiency parameters.

Note that our tradeoff parameters (total and brown power consumption) depend on the number
of connected data center nodes to the FE node. In other words, connecting to a different number of
data centers will result in a different amount of total and brown power consumption. After connecting to data center nodes, the FE node becomes and remains inactive, and selected data center nodes become active. Note that each data
center node is represented by a pair of incoming and outgoing nodes, i.e., $x_{in}$ and $x_{out}$, connected by a directional edge whose capacity is the
maximum number of requests that the data center can handle
by the deadline, i.e., its UB. Finally, when the deadline comes, the data
collector node becomes active and connects to the data center
nodes to receive the processed requests. The edges that connect
from the data center nodes to the data collector node are
assumed to have infinite capacity, i.e., users have access to
all the processed requests.

\subsection{Total-Brown Tardeoff}
\begin{figure}
\center
\includegraphics[width=9cm,height=8cm]{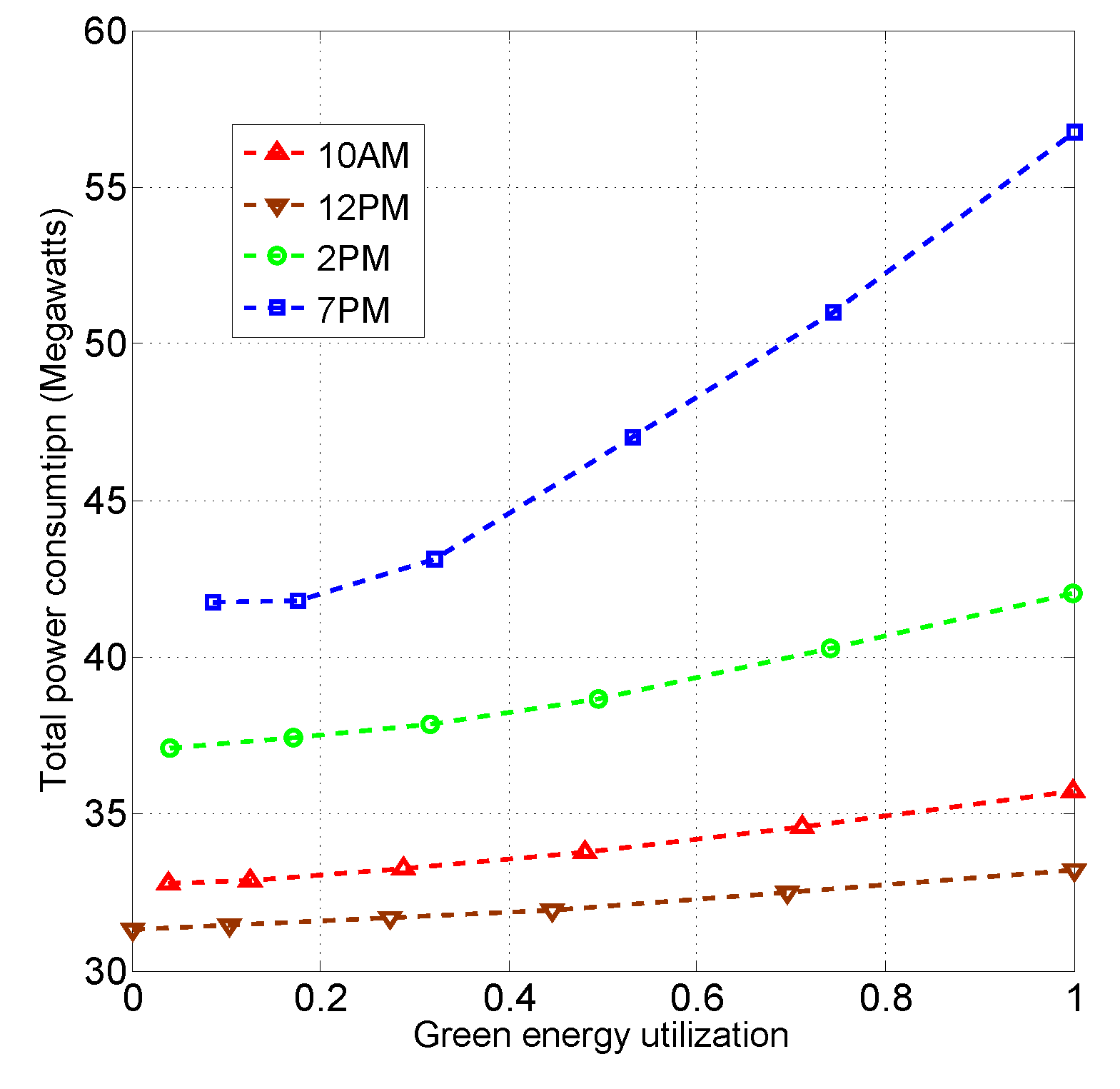}
\caption{Green power utilization-total power consumption tradeoff curves at different hours of day.}
\label{fig:6}
\end{figure}

The proposed information flow graph model can be used to capture the whole trade-off region between
the total and brown power consumption. The main idea is that the FE has to distribute the workload to data centers such that
the capacity of the FE-data collector minimum cut
is larger than or equal to the total number of incoming requests at FE. If so, the data collector node can ensure receiving all the processed requests by the deadline, and the workload distribution strategy can meet the SLA requirements.
As a numerical example, we consider $k=6$ data centers, each integrated with a
wind farm as a renewable power source. Our simulation data are based on the trends of wind power and the total workload shown in Figs.~\ref{fig:3} and~\ref{fig:4}, respectively. Fig.~\ref{fig:6} shows the tradeoff curves between the total power consumption and green energy utilization at some sample
hours of the day. For example, at time 10AM, if we utilize all available green energy at 6 data centers, i.e., when the green energy utilization is 1, more than 35Megawatts of power is consumed totally. The tradeoff curves in this figure confirm that we can increase the green energy utilization by increasing the total power consumption.

\section{Summary}
Owing to the cost and scale, a few number of the Internet scale data centers are dispersed over the globe. To this end, geographical load balancing schemes are designed to leverage this geo-dipersion by efficient utilization of global resources such as price diversity
in the electricity markets or locational diversity in the renewable power generation, i.e., distributing the requests to the locations with cheaper price of electricity or higher green power generation.

Because of the different costs and different environmental impacts of the renewable energy and brown energy, geographical load balancing approaches can be further benefitted by the idea of green
workload and green service rate, versus brown workload and
brown service rate, respectively. In fact, the concept
of decomposing the workload to the workloads served by green and brown energy facilitates the separation of
green energy utilization maximization and brown energy cost minimization problems.

 On the other hand, utilizing price and green energy diversities via geographical load balancing can increase the total power consumption due to the fact that different data centers have different servers with different service capabilities, and also a request sent to different data centers experiences different network delays.
  Therefore, there is a tradeoff between total power consumption and green energy utilization or cost of the electricity. Such a tradeoff can be captured by modeling the geo-dispersed data centers with an information flow graph and more importantly, by defining a service efficiency parameter based on an M/GI/1 Processor Sharing (PS) queue to take into consideration of the network delay between FE servers and data centers.

Extending the green versus brown idea and information flow graph based model by considering multiple workload distribution centers as the geographically concentrated sources of requests, the availability of energy storage devices at the data centers that may introduce new challenges like some battery related constraints, and elaborating on machine learning based energy prediction techniques will be an interesting future pursuit.
\bibliographystyle{IEEE}
\bibliography{refs}
\end{spacing}
\end{document}